Jeff Jones, Richard Mayne, Andrew Adamatzky
Centre for Unconventional Computing, University of the West of England,
Coldharbour Lane, Brisol, BS16 1QY, UK.

# Representation of Shape Mediated by Environmental Stimuli in *Physarum polycephalum* and a Multi-agent Model

November 17, 2015

**Abstract**

The slime mould *Physarum polycephalum* is known to construct protoplasmic transport networks which approximate proximity graphs by foraging for nutrients during its plasmodial life cycle stage. In these networks, nodes are represented by nutrients and edges are represented by protoplasmic tubes. These networks have been shown to be efficient in terms of length and resilience of the overall network to random damage. However relatively little research has been performed in the potential for *Physarum* transport networks to approximate the overall shape of a dataset. In this paper we distinguish between connectivity and shape of a planar point dataset and demonstrate, using scoping experiments with plasmodia of *P. polycephalum* and a multi-agent model of the organism, how we can generate representations of the external and internal shapes of a set of points. As with proximity graphs formed by *P. polycephalum*, the behaviour of the plasmodium (real and model) is mediated by environmental stimuli. We further explore potential morphological computation approaches with the multi-agent model, presenting methods which approximate the Convex Hull and the Concave Hull. We demonstrate how a growth parameter in the model can be used to transition between Convex and Concave Hulls. These results suggest novel mechanisms of morphological computation mediated by environmental stimuli.

## 1 Introduction

Slime mould *Physarum polycephalum* is a single-celled organism which is capable of remarkable biological and computational feats, despite possessing no nervous system, skeleton or organised musculature. The study of the computational potential of the *Physarum* plasmodium was initiated by Nakagaki et al. [1] who found that the plasmodium could solve simple maze puzzles. This research has been extended and the plasmodium has demonstrated its performance in, for example, path planning and plane division problems [2], spanning trees and proximity graphs [3], [4], simple memory effects [5], the implementation of individual logic gates [6] and *Physarum* inspired models of simple adding circuits [7].

The plasmodium of slime mould is amorphous in shape, ranging from the microscopic scale to up to over a square metre in size. It is a giant multinucleate syncytium comprised of a sponge-like actomyosin complex co-occurring in two physical phases. The gel phase is a dense matrix subject to spontaneous contraction and relaxation, under the influence of changing concentrations of intracellular chemicals. The protoplasmic sol phase is transported through the plasmodium by the force generated by the oscillatory contractions within the gel matrix. Protoplasmic flux, and thus the behaviour of the organism, is affected by local changes in temperature, space availability, chemoattractant stimuli and illumination [8], [9], [10], [11]. The *Physarum* plasmodium can thus be regarded as a complex functional material capable of both sensory and motor behaviour. Indeed *Physarum* has been described as a membrane bound reaction-diffusion system in reference to both the complex interactions within the plasmodium and the rich computational potential afforded by its material properties [12].

Computational geometry problems tackle the grouping or partitioning of points in the plane or in higher dimensions. Because of the lack of supportive tissue, the plasmodium typically extends along the space of the surface on which it lives, and *Physarum* may be considered as a 2D organism and its nutrient sources can be considered as a coarse representation of points on the plane. Conversely, the networks formed by growth and adaptation of the organism can be considered as edges in the plane. The efficiency of proximity graphs formed by *Physarum* is a trade-off between minimum distance (or minimum amount of material) and resilience to random disconnections [13]. However, although the networks connect all of the nutrient sources, they do not group them, or provide a representation of the space or shape in which they reside.

Relatively little research has been performed in assessing the behaviour of *Physarum* on representing the area or shape of a set of points. The plasmodium was previously used to perform division of the plane in approximations of Voronoi diagrams. The Voronoi diagram of a set of $n$ points in the plane is the subdivision of the plane into $n$ cells so that every location within each cell is closest to the generating point within that cell. Conversely the bisectors forming the diagram are equidistant from the points between them. Two different methods have been proposed. The first method used avoidance of nodes represented by repellent sources [2, 14] and the resultant pattern of plasmodial veins approximated the Voronoi bisectors. The second method utilised the merging of growth fronts of individual plasmodia to represent the Voronoi bisectors [15] at regions where fusion of the plasmodia occurred. A multi-agent model of *Physarum* was used to model both methods of Voronoi diagram approximation and was found to generate unusual hybrid graphs which combined plane division (grouping and division of points) with internally minimal connections between the points [16].

Although the Voronoi diagram can be considered as representing area (dividing a set of points), it does not represent the overall shape or border of a set of points. A method of representing the overall shape was described in [17] using stimuli which had a long-range attractant effect and a short-range repellent effect. By inoculating the plasmodium away from the set of points the organism grew towards the attractants but then, repelled at short range, traversed the periphery of the stimuli, approximating the Concave Hull of the point set.

Why does *Physarum* not naturally represent the shape of its environment? One reason may be that the organism appears to behave in a manner which initially optimises (maximises) area exploration and which later adapts its network by optimising (minimising) network distance and network resiliency to damage. This behaviour has previously been described as a trade-off between exploration and exploitation [18]. When inoculating the plasmodium in a shape pattern, the organism will quickly form networks, breaking up the solid pattern. This is an efficient strategy in terms of minimising material and energy resources but, from a computational perspective, is not useful if we require a shape representation. By using a multi-agent model of *Physarum* which behaves as an adaptive virtual material is is possible to slow the adaptation of the (virtual) plasmodium so that, as it adapts, it retains its solid shape. This was used in a simple method so approximate a combinatorial optimisation problem by shrinkage [19]. During this shrinkage process a transition continuum was seen from the complete coverage of the data set (the initial inoculation pattern) down to the Steiner Minimum Tree. This suggests the possibility of using morphological adaptation to compute the area occupied by, and the general shape of, a set of points.

In this paper we examine potential mechanisms, mediated by environmental stimuli, to represent the area and shape of a set of points using *Physarum* and its multi-agent model. We begin by assessing the possibility of confining the plasmodium to represent shape using attractants and light illumination. We reproduce these results in a multi-agent model and then extend the modelling approach to examine different methods of approximating the Convex Hull and the Concave Hull. We devise a parameter which can be used to control the concavity of a growing model plasmodium.

## 2 Experimental Results

It is known that the growth of *P. polycephalum* is affected by stimuli within its environment. These stimuli include the presence and distribution pattern of chemo-attractants within its local environment (to which the organism grows towards) and exposure to hazards such as light irradiation (which the organism avoids). The growth patterns of *P. polycephalum* are, however, difficult to control with any accuracy. Can attractants and light irradiation be used to 'persuade' the organism to conform to a particular shape?

In Fig. 1, a *P. polycephalum* plasmodium was inoculated onto a 2% non-nutrient agar plate upon which an array of chemo-attractants (oat flakes) was arranged in the shape of the letter 'H' (full details of plasmodial culture technique are included in appendix 1). On the lid of the plate, an H-shaped cardboard mask was present that completely covered the oats. The plasmodium was then left to propagate for approximately 3 days in a light-proof box in the presence of a bright white light (a 7W array of 48 5500K 'daylight white' LEDs, 156cd (Lighting Ever, UK)), causing all portions of the plate except the area under the cardboard mask to be irradiated. The organism was observed to grow towards and subsequently connect the oat flakes: the mask mostly was found to prevent the organism from foraging into illuminated areas and the subsequent plasmodial network approximated the 'H' shape (Fig. 1h).

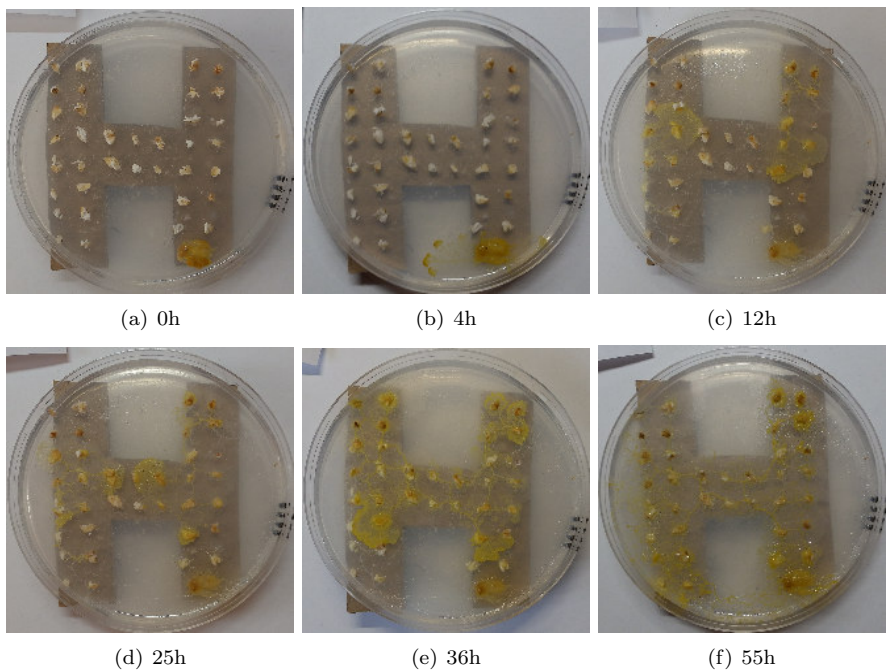

(a) 0h  (b) 4h  (c) 12h
(d) 25h  (e) 36h  (f) 55h

Figure 1: Environmentally mediated confinement of plasmodium for shape representation. a) inoculation of *P. polycephalum* (bottom right) on an agar plate loaded with oat flakes patterned in letter 'H' configuration. Cardboard shape indicates region masked from light illumination, b-f) propagation of plasmodium connects oat flakes whilst tending to avoid illuminated regions outside the mask.

In the corresponding control experiment, illumination was removed to assess whether the attractant stimuli alone could confine the plasmodium to the shape. As Fig. 2 demonstrates, the lack of illumination resulted in no representation of the 'H' pattern, even though the oat flakes retained this pattern. The final configuration shows almost uniform coverage of the arena, regardless of the oat flake configuration (Fig. 2h, suggesting that the hazardous stimulus is important for shape representation).

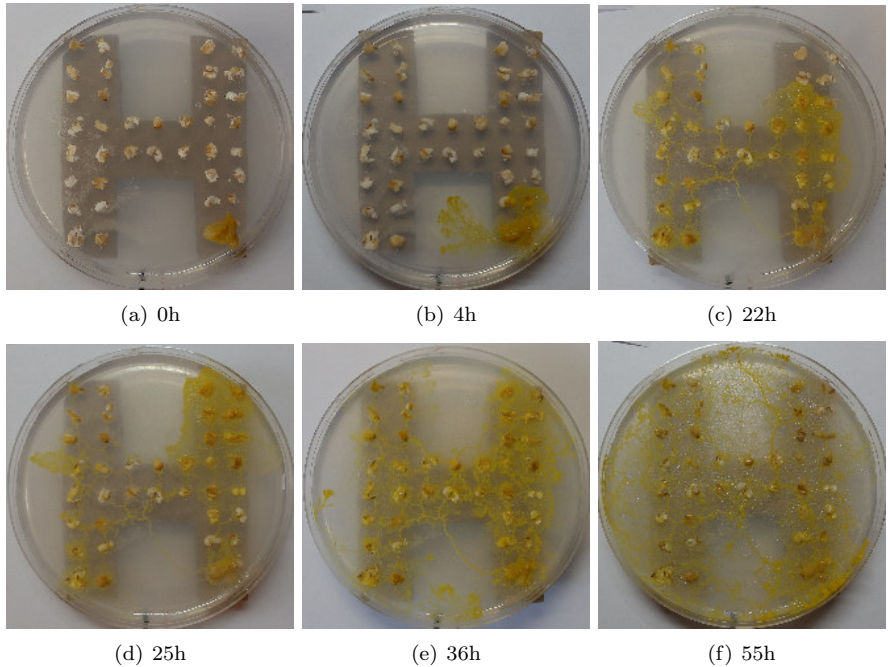

Figure 2: Removal of illumination stimuli prevents confinement to shape. a) inoculation of *P. polycephalum* on oat flakes patterned in letter 'H' configuration. Control experiment with no illumination stimulus, b-f) propagation of plasmodium connects oat flakes but does not avoid regions outside the cardboard mask.

In further studies using an oat/mask configuration in the shape of the letter 'C' (Fig. 3), it was found that, once again, experimental plasmodia approximate the shape of the nutrient array when a corresponding protective mask is present. Having fully occupied the space under the mask, however, the plasmodium was observed to extend a growth front to connect opposite sides of the 'C' shape. This can be seen in the superimposed and enhanced collective representation of the plasmodial pattern shown in Fig. 3g (for details of image enhancement, see appendix 2). The final pattern of the network approximates the convex hull for this point set (Fig. 3h).

It was found that the presence of an illumination mask was necessary to confine the plasmodium within a shape. How important is the contribution of the attractant array in comparison to the mask? To answer this, the 'C' shape was projected onto plates with an illumination mask but presented the array of attractant stimuli not in the shape, but as a regularly spaced ($10 \times 10mm$ apart) 2D array (Fig. 4a). When inoculated within this array, the plasmodium was found to occasionally migrate towards attractants outside the mask (Fig. 4c, for example) but the majority of the plasmodial network was confined within the unilluminated region. The importance of the illumination mask is again emphasised by the uniform coverage of the 2D attractant array when the illumination mask pattern is not presented (Fig. 4h). The results presented here are highly repeatable.

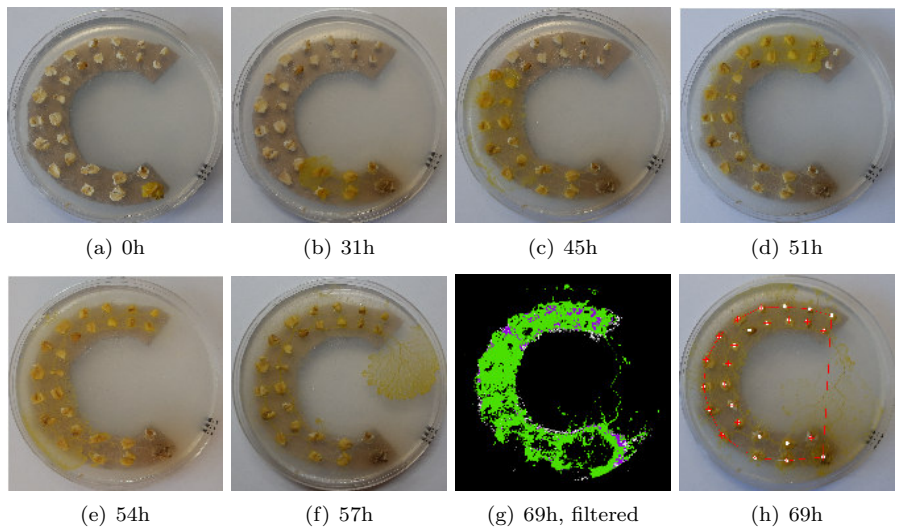

| (a) 0h | (b) 31h | (c) 45h | (d) 51h |
| (e) 54h | (f) 57h | (g) 69h, filtered | (h) 69h |

Figure 3: Environmentally mediated confinement of plasmodium for shape representation. a) inoculation of *P. polycephalum* on oat flakes patterned in letter 'C' configuration. Cardboard shape indicates region masked from light illumination, b-e) propagation of plasmodium connects oat flakes whilst tending to avoid illuminated regions outside the mask, f-h) after reaching the end of the shape a protoplasmic tube is extended to connect the opposite side of the shape, approximating the Convex Hull of the shape (overlaid), g) view of superimposed plasmodial network subjected to image enhancement (see text for details) indicates that the majority of the network is confined by the illumination mask within the shape.

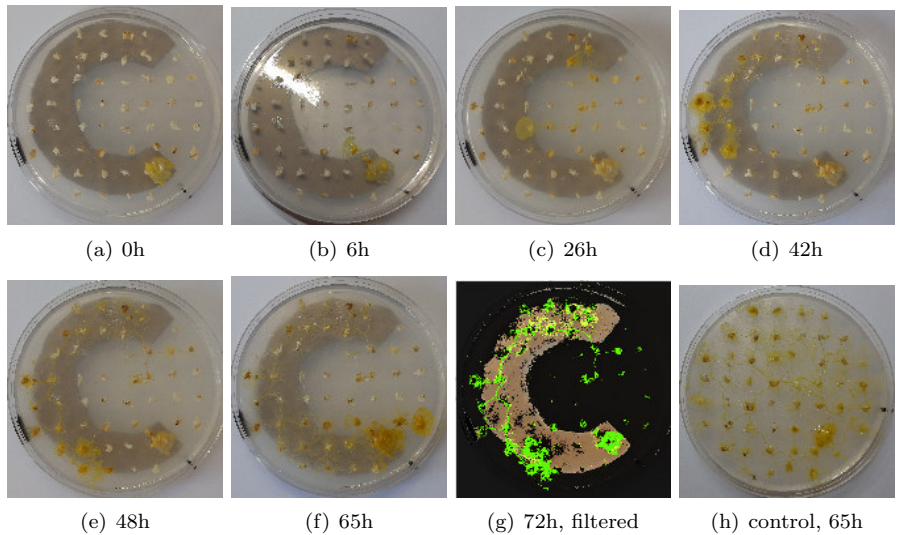

(a) 0h　　(b) 6h　　(c) 26h　　(d) 42h

(e) 48h　　(f) 65h　　(g) 72h, filtered　　(h) control, 65h

Figure 4: Environmentally mediated confinement of plasmodium for shape representation. a) inoculation of *P. polycephalum* on oat flakes arranged in a regular 2D array. Cardboard shape indicates shaped region masked from light illumination, b-f) propagation of plasmodium connects oat flakes whilst tending to avoid illuminated regions outside the mask, g) superimposed plasmodium networks subjected to image enhancement (see text for details) indicating that the majority of the the network is confined within the boundaries of the mask, h) control experiment without illumination shows uniform growth within the array and no representation of shape.

## 3  Modelling Results

We used the multi-agent approach introduced in [20]. Agents sense the concentration of a hypothetical 'chemical' in a 2D lattice, orient themselves towards the locally strongest source and deposit the same chemical during forward movement. The agent population spontaneously forms emergent transport networks which undergo complex evolution, exhibiting minimisation and cohesion effects. The dynamical network patterns were found to reproduce a wide range of Turing-type reaction-diffusion patterning [21]. External stimuli by nutrients and repellents are represented by projecting positively weighted and negatively weighted values respectively into the lattice and the network evolution is constrained by the distribution of nutrients and repellents. Network evolution is affected by nutrient distribution, nutrient concentration and repellent placement. A full description of the model and parameters is given in appendix 3.

### 3.1 Experimental Validation

The model was initially used to replicate the experimental results. We patterned a set of attractant stimuli in the lattice in the 'H' shape as used in the previous experiment and inoculated a small population at a single stimulus site. The model was run with simulated light irradiation and again without. The results showed that the simulated light illumination mask was necessary for the model to be mostly confined within the H pattern (Fig. 5a-d). Without the illumination mask the population was not confined to the shape and grew a transport network which occupied the entire arena (Fig. 5e-h).

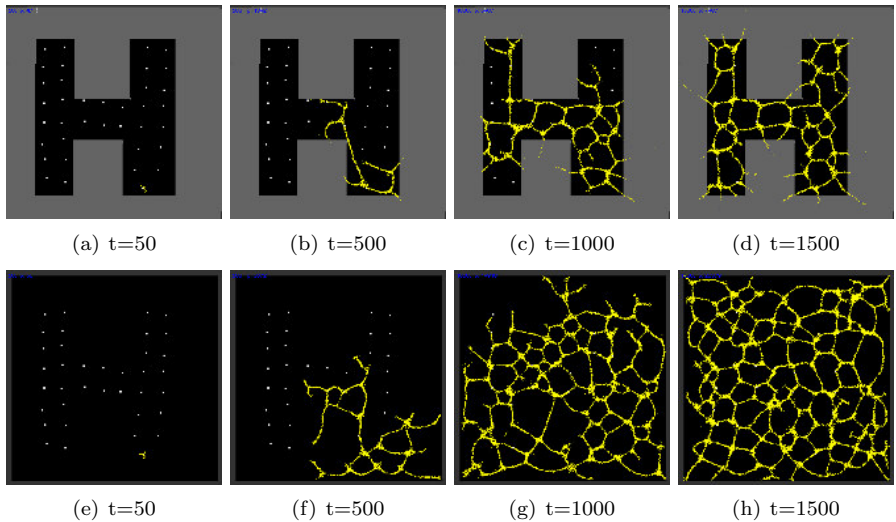

(a) t=50  (b) t=500  (c) t=1000  (d) t=1500

(e) t=50  (f) t=500  (g) t=1000  (h) t=1500

Figure 5: Approximation of response to 'H' pattern of attractants in multi-agent model with and without simulated light irradiation. a-d) growth of model plasmodium with simulated light irradiation outside masking area. model plasmodium does not venture outside the masked area, e-h) growth of model plasmodium without light irradiation mask shows unconstrained growth outside the pattern of attractants.

These model results reproduced the experimental findings. However, the model differs from the real plasmodium in that it is not affected by adhesion to the slime matrix, a polysaccharide-rich protective extracellular 'sheath'. This allows greater flexibility when using the *P. polycephalum* plasmodium as an inspiration to develop spatially represented unconventional computation methods. Using such approaches we have been able to demonstrate morphological computation of combinatorial optimisation problems [19], spline curves [22] and the centre-of-mass of a shape [23]. It should be noted that such methods may not be directly implementable in real plasmodium due to practical limitations and the unpredictable behaviour of the organism. However the methods do follow the same basic qualities, namely simple material behaviour, morphological adaptation and distributed embedded computation.

## 3.2 Convex Hull by Material Shrinkage Around Attractants

We begin by devising approaches to approximate the Convex Hull. The Convex Hull of a set of points is the smallest convex polygon enclosing the set, where all points are on the boundary or interior of the polygon (Fig. 6a). Classical algorithms to generate Convex Hulls are often inspired by intuitively inspired methods, such as shrink wrapping an elastic band around the set of points, or rotating calipers around the set of points [24, 25]. Is it possible to approximate the Convex Hull using emergent transport networks by mimicking a physically inspired method? To achieve this we initialised a circular ring of virtual plasmodium *outside* the set of points (Fig. 6b). Because of the innate minimising behaviour of the particle networks the population thus represented a ring of deformable elastic material.

This bounding 'band' then automatically shrinks to encompass the outer region of the set of points. The minimising properties of the paths ensure that the edges of the Hull are straight and convex. There are some practical limitations of this approach. Firstly, the bounds of the set of points must be known in advance, which is not always the case in certain Convex Hull problems. Secondly, points which are inside the final Hull, but close to the 'band' (for example near the top edge in Fig. 6c) may, via diffusion of their projected attractant, attract the band inwards, forming a concavity. This possibility may be avoided by restricting the nodes to project stimuli only when they have been directly contacted by particles comprising the shrinking band. One benefit of this is that the nodes which are actually part of the final Hull are highlighted (Fig. 6e, the larger nodes).

## 3.3 Convex Hull by Material Shrinkage Around Repellents

Alternatively, to avoid the potential of attraction to nodes within the Hull boundary, it is possible to have the 'band' shrink around the array of points which are actually *repulsive* to the particles comprising the band. This is achieved by projecting a repellent stimulus (for example, a negative value into the lattice) at the nutrient node locations. The band will still shrink to envelope the nodes but — because of the repulsion effect — will not actually contact the nodes. This generates a Convex Hull which encompasses the nodes but does not directly touch them (Fig. 7) and results in a Hull which slightly overlaps the original dataset.

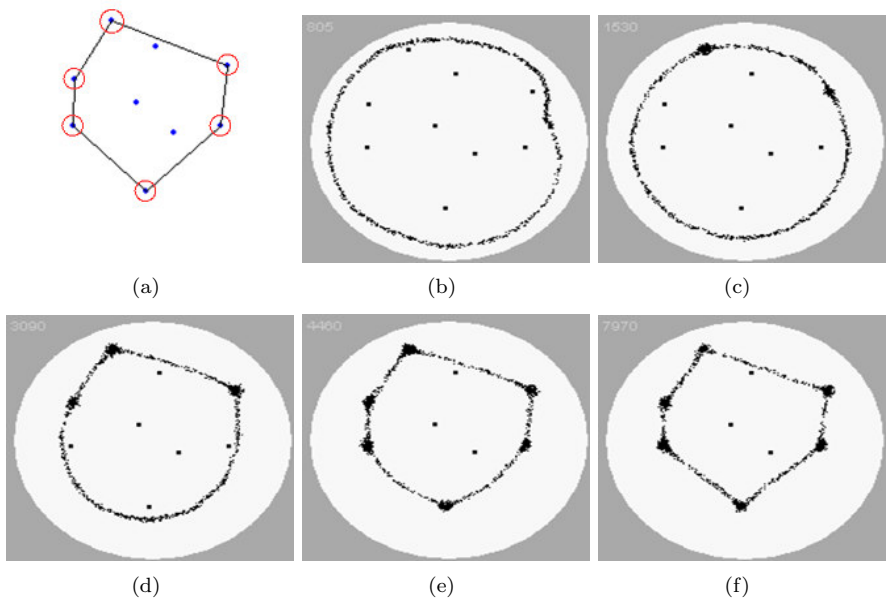

Figure 6: Approximation of Convex Hull by shrinking band of virtual plasmodium. a) original data set with Convex Hull (edges). Nodes which are part of the Convex Hull are circled, b-e) A circular band of virtual plasmodium initialised outside the region of points and shrinks. In this example nodes only emanate nutrients when touched by virtual plasmodia (see text), f) bounding points of final Convex Hull are indicated by larger nodes.

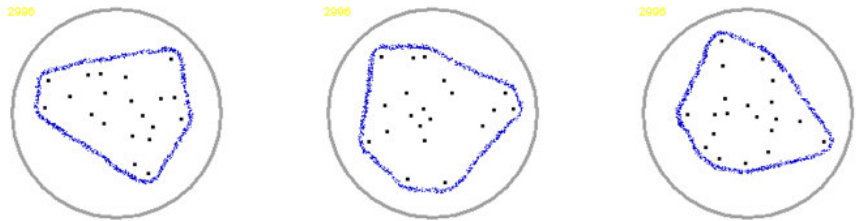

Figure 7: Convex Hull via shrinkage around repellent stimuli. Three separate examples are shown. A band of virtual plasmodium shrinks around the set of points to approximate the Convex Hull. Note a small peripheral region is indicated because of the repulsive region.

### 3.4 Convex Hull by Self-organisation

If the boundary of the Hull points is not known in advance then it is possible to utilise a method which employs both self-organisation and repulsion to approximate the Hull, as shown in Fig. 8. In this approach the particle population is initialised at random locations within the lattice (both outside and inside the set of points). The particles are repelled by the repellent nodes and move away from these regions. If a particle touches a node it is annihilated and randomly initialised to a new blank part of the lattice. Over time, the inner region of the lattice becomes depleted of particles, but in contrast the region outside the set of point (which is further away from the repulsive nodes) becomes more populous. The increasing strength of the emerging Convex Hull trail outside the dataset attracts particles from inside the dataset (because the deposited 'ring' of flux is higher in concentration than the inner region, due to the increased number of particles) and the particles are drawn out into this ring. The natural contraction of the outer ring approximates the final Convex Hull.

### 3.5 Representing the Shape of a Set of Points

The area occupied by, or the 'shape' of, a set of points is not as simple to define as its Convex Hull. It is commonly defined in Geographical Information Systems (GIS) as the Concave Hull, the minimum region (or *footprint* [26]) occupied by a set of points, which cannot, in some cases, be represented correctly by the Convex Hull [27]. For example, a set of points arranged to form the capital letter 'C' would not be correctly represented by the Convex Hull because the gap in the letter would be closed (see Fig. 10a).

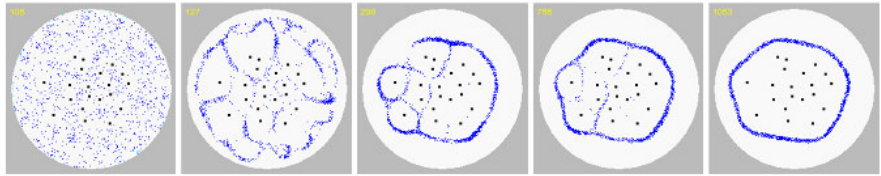

Figure 8: Convex Hull via self-organisation within repulsive field. Particle population is initialised randomly in the arena and is repulsed by nodes. Convex Hull emerges at the border and internal connections gradually weaken.

Attempts to formalise concave bounding representations of a point set were suggested by Edelsbrunner et al. in the definition of $\alpha$-shapes [28]. The $\alpha$-shape of a set of points, $P$, is an intersection of the complement of all closed discs of radius $1/\alpha$ that includes no points of $P$. An $\alpha$-shape is a Convex Hull when $\alpha \to \infty$ (Fig. 9a). When decreasing $\alpha$, the shapes may shrink, develop holes and become disconnected (Fig. 9b-d), collapsing to $P$ when $\alpha \to 0$. A Concave Hull is non-convex polygon representing area occupied by $P$. A Concave Hull is a connected $\alpha$-shape without holes.

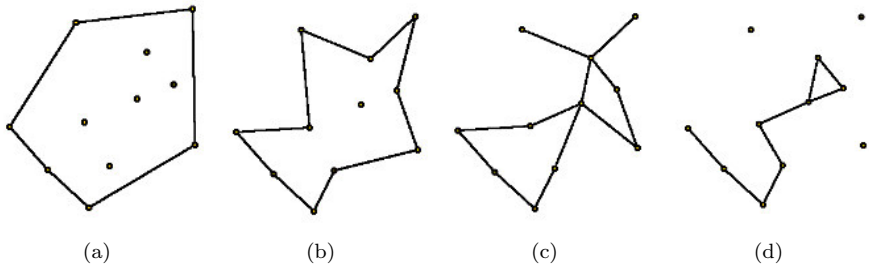

(a)     (b)     (c)     (d)

Figure 9: Examples of $\alpha$-shape of a set of points as $\alpha$ decreases. Note the limitations of this approach as the shapes can contain cycles (c) or become disconnected from the data points (d).

## 3.6 Approximation of the Concave Hull by Shrinkage

The virtual plasmodium approximates the Concave Hull via its innate morphological adaptation as the population size is slowly reduced. A slow reduction in population size prevents hole defects forming in the material which would result in cyclic networks instead of the desired solid shape. The reduction in population size may be implemented by either randomly reducing particles at a low probability rate or by adjusting the growth and shrinkage parameters to bias adaptation towards shrinkage whilst maintaining network connectivity.

In the examples shown below the virtual plasmodium is initialised as a large population (a solid mass) within the confines of a Convex Hull (calculated using the classical algorithmic method) of a set of points (Fig. 10b). By slowly reducing the population size (by biasing the parameters towards shrinkage), the virtual plasmodium adapts its shape as it shrinks. Retention of the mass of particles to the nodes is ensured by chemoattractant projection and as the the population continues to reduce, the shape outlined by the population becomes increasingly concave (Fig. 10c-f).

The graph of changing population size as the virtual plasmodium adapts (Fig. 11) shows that the population stabilises as the concave shape is adopted. If varying degrees of concavity are required, the current population size as a fraction of the original size, or alternatively the rate of population decline, could possibly be used as a simple parameter to tune the desired concavity.

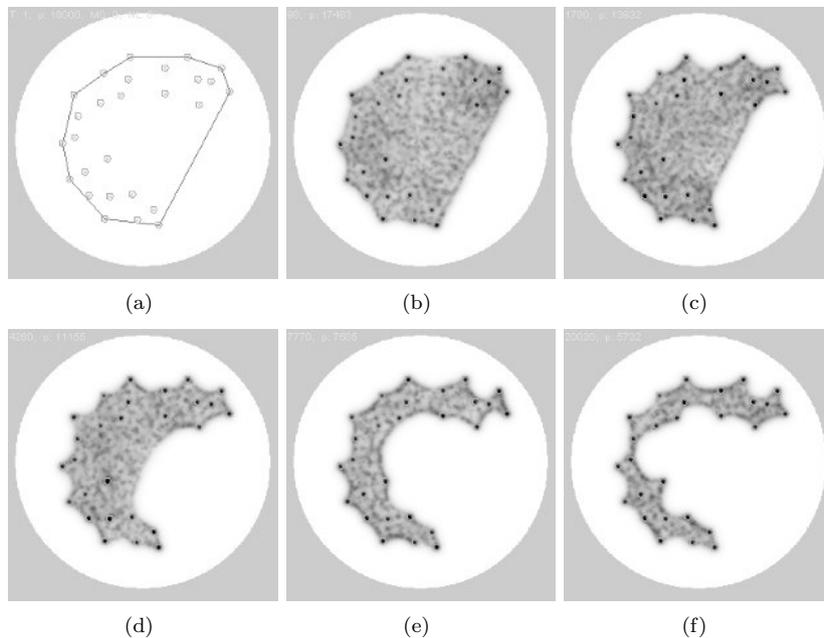

Figure 10: Concave Hull by uniform shrinkage of the virtual plasmodium. (a) Set of points approximating the shape of letter 'C' cannot be intuitively represented by Convex Hull, (b-f) Approximation of concave hull by gradual shrinkage of the virtual plasmodium, $p$=18,000, $SA$ 60°, $RA$ 60°, $SO$ 7.

If the shrinkage of the initial Convex Hull were to continue beyond the Concave Hull the area would shrink until a network representation (approximating the Steiner tree) is formed. The shrinkage of the agent population thus represents the transition between area coverage and network distance.

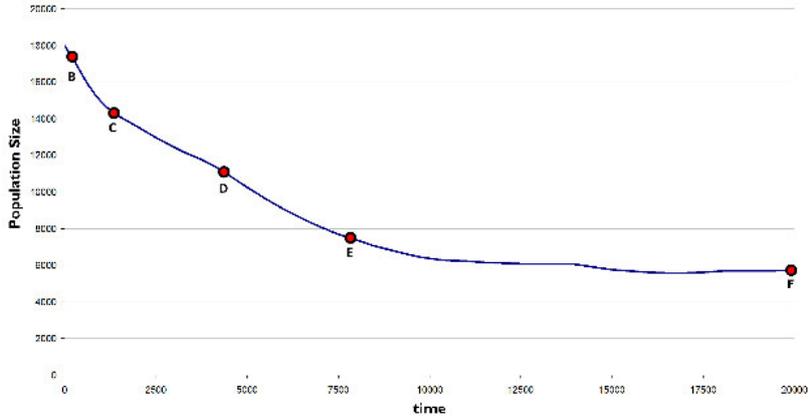

Figure 11: Decrease in population size as concave shape formed from a Convex Hull of the point set in Fig. 10a. Plot shows population size over time. Letters and circles B-E represent population levels in the corresponding images of Fig. 10(b–e).

## 3.7 Approximating the Concave Hull by Growth

The shrinkage of a solid mass of virtual plasmodium cannot construct $\alpha$-shapes, shapes with vacant regions within them, for example as with the letter 'A'. However, by initialising the a smaller population size at the node sites themselves, the individual fragments of 'plasmodium' grow and fuse together when each fragment senses the attractant deposited by a neighbouring fragment, eventually recovering the general shape of the letter (Fig. 12,a-d). Further increasing the population size (manually or by biasing growth/shrinkage parameters) results in removal of the internal space and transition from an $\alpha$-shape to a solid Concave Hull (Fig. 12,e-f).

One limitation with this approach is that it cannot guarantee that all sites will fuse. For example, if one node is a significant distance from all other sites it will not sense the stimulus from more distant sites. This node will thus not fuse with the remaining masses, resulting in two separate shapes. This limitation can be overcome by ensuring that the initial inoculation sites are connected in some way. A suitable candidate pattern is the Minimum Spanning Tree (MST) structure of the data points (Fig. 13a and b). This structure guarantees connectivity between all points and also does not possess any cyclic regions. By inoculating the model plasmodium on the MST pattern and biasing the growth/shrinkage parameters towards growth, the model then 'inflates' the MST (Fig. 13c-i) and automatically halts its growth (maintaining a constant population size) as a Concave Hull is approximated (Fig. 14). To visualise the classical Concave Hull edges from this pattern we can use the approach described in [19] and traverse the perimeter of the pattern, constructing the Concave Hull by adding nodes which are located on the periphery of the shape,

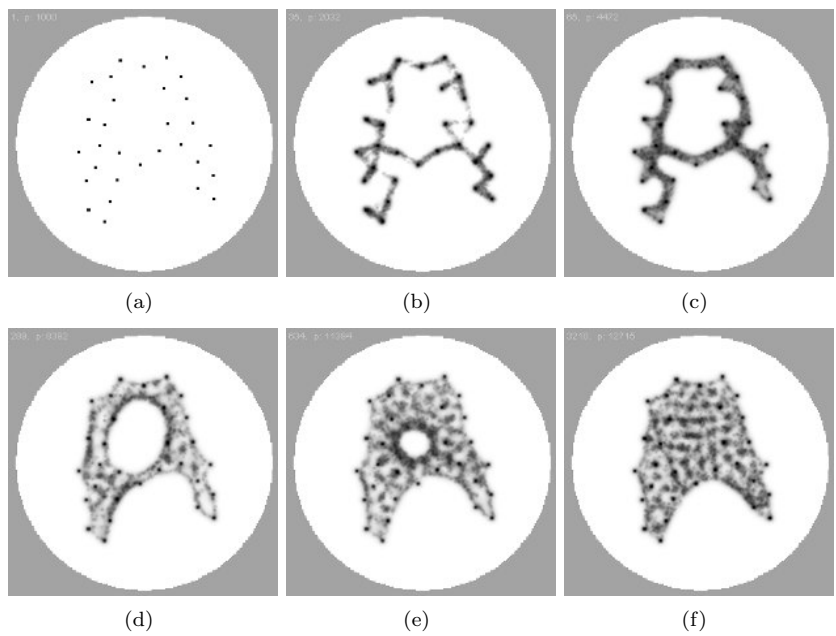

Figure 12: Alternate method of generating $\alpha$-shape and Concave Hull by merging regions.

yielding the classical (straight lines) structure of the Concave Hull (Fig. 13i, edges).

## 3.8 Transformation Between Convex and Concave Hull

By growing the population from the MST a representation of the Concave Hull is generated. Can we 'tune' the shape of the final blob pattern? In Fig. 15 we assess this on a simple test structure, a set of points arranged in a square. After inoculation on the MST the population grows and at 5000 steps the blob fills the region between the bounding points (in this square pattern the Concave Hull is the same as the Convex Hull). Galton demonstrated that the an optimal descriptor for the 'shape' of a set of points is a conflicting trade-off between minimising area of the shape and minimising the perimeter of the shape [29]. Can we utilise a single parameter of the multi-agent model to transition between these two competing objectives?

By adjusting the value of the $G_{max}$ growth parameter it was found that the concavity of the blob could be adjusted. Fig. 16 shows the effect of different values of $G_{max}$ in five different experiments. As $G_{max}$ increased, the concavity reduced until at $G_{max} = 25$ the final blob was fully convex (maximising the area and minimising the perimeter). At higher $G_{max}$ values the growth was not constrained by the stimuli from the point sources, causing uncontrolled growth patterns (Fig. 16e). We also found that reducing the value of the $G_{max}$ parameter during an experimental run dynamically reduced the size and area of the blob, transforming convex shapes back into concave shapes (minimising the area but increasing the perimeter).

# 4 Conclusions

The results in this paper demonstrate that it may be possible to utilise slime mould *Physarum polycephalum* to approximate the external and internal shape of a set of points. We showed experimentally how this can be achieved using chemo-attractant stimuli and masking by light illumination. We reproduced these results in a multi-agent model of slime mould. We then extended the multi-agent approach to investigate kinaesthetically inspired approaches to problems in which a representation of shape is needed. This is not seen in the behaviour of *P. polycephalum* itself, due to the spontaneous formation of networks and its relatively unpredictable behaviour. However it is possible to bias the pattern formation mechanisms of the model to generate different shrinkage behaviours. In the case of the Convex Hull we approximated the intuitive 'band' method but were also able to generate novel methods based on repulsion and self-organisation. The Concave Hull was approximated by shrinkage from the Convex Hull and also by growth. Growth-based approximation of the Concave Hull was initially performed by fusion of individual virtual plasmodia inoculated on point sources but this can generate disconnected shapes. By inoculating the population on the MST it was possible to grow fully connected

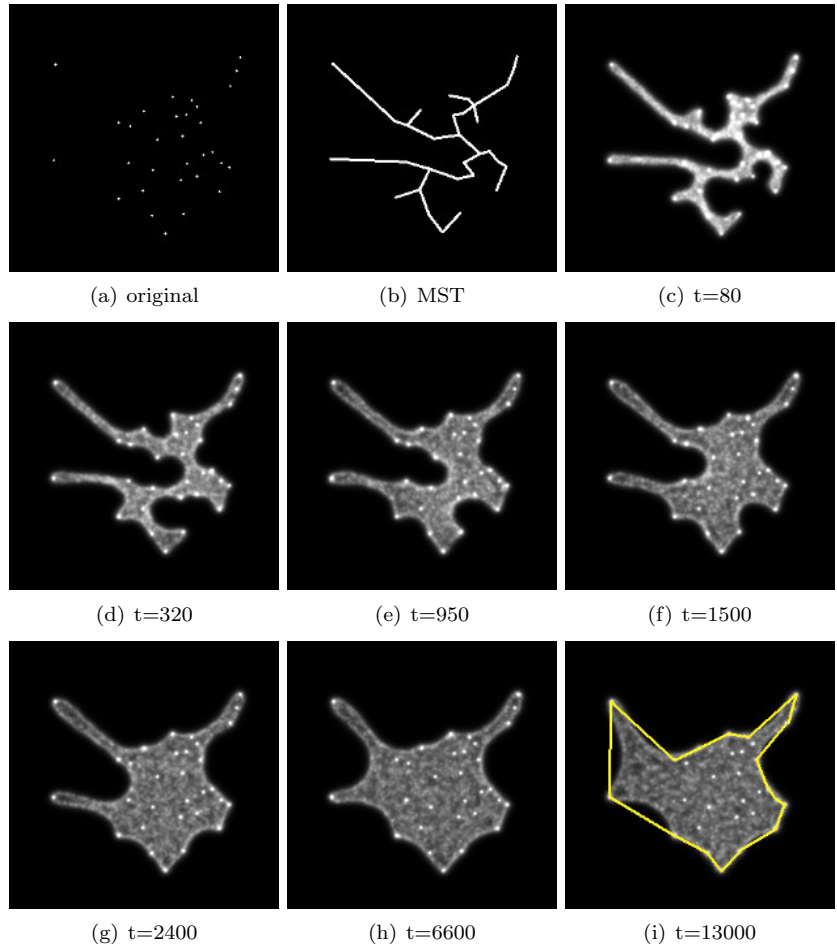

Figure 13: Growth of Concave Hull from Minimum Spanning Tree. a) points representing the locations of major cities in People's Republic of China, b) Minimum Spanning Tree of points connects all points without cycles, c-i) after inoculating the virtual plasmodium on the Minimum Spanning Tree the virtual plasmodium grows to approximate the Concave Hull, stabilising its growth automatically (overlaid edges show classical Concave Hull).

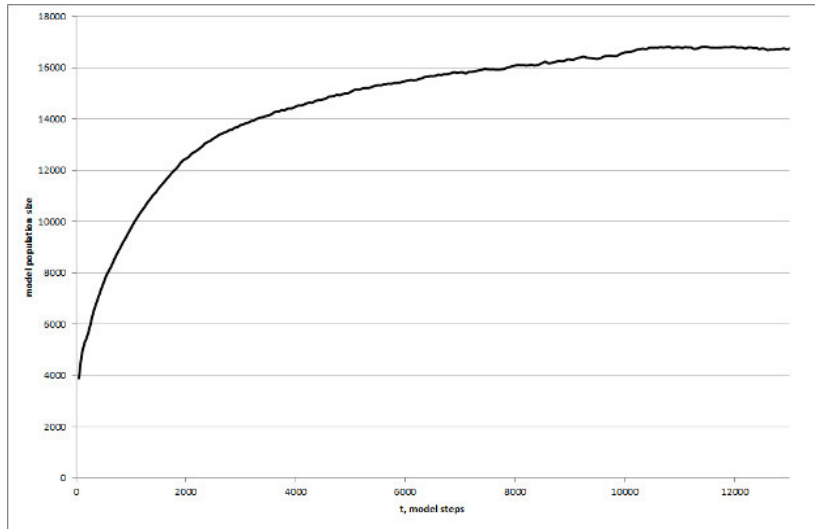

Figure 14: Increase and stabilisation of population size as concave hull grows from inoculation of the model plasmodium on the MST. Plot shows population size over time.

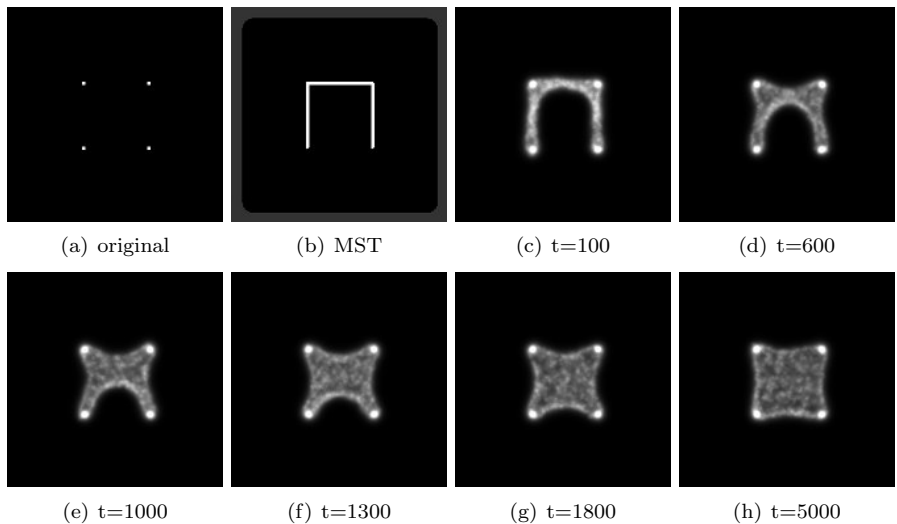

(a) original  (b) MST  (c) t=100  (d) t=600

(e) t=1000  (f) t=1300  (g) t=1800  (h) t=5000

Figure 15: Tuning concavity of growth from Minimum Spanning Tree. a) 4 points arranged in square pattern, b) Minimum Spanning Tree of square points used as inoculation sites, c-h) after inoculation on the Minimum Spanning Tree the virtual plasmodium grows to approximate the Concave Hull.

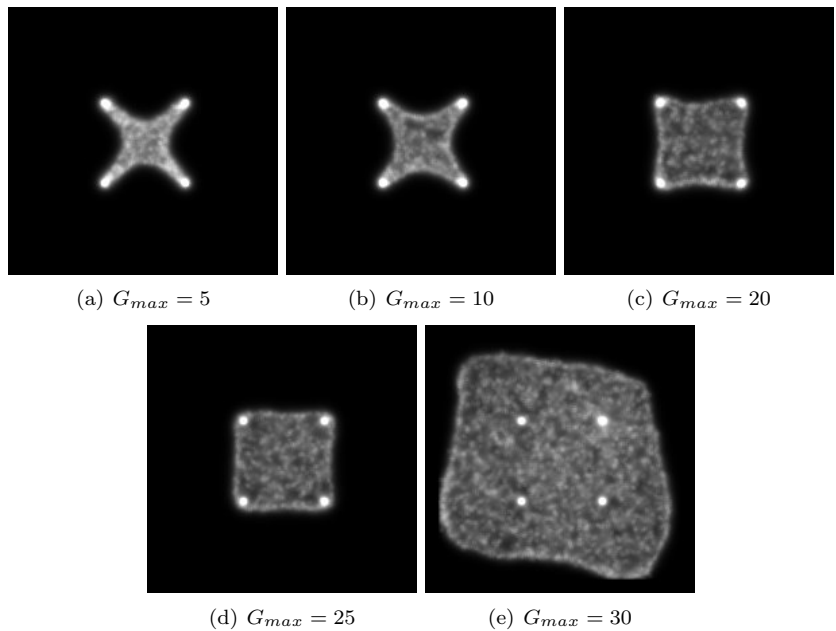

Figure 16: Increasing $G_{max}$ growth parameter reduces blob concavity. Separate experiments run for 5000 steps, except in the case of 'e' which was halted at 3000 steps due to unconstrained growth. a) $G_{max} = 5$, b) $G_{max} = 10$, c) $G_{max} = 20$, d) $G_{max} = 25$, e) $G_{max} = 30$.

Concave Hulls and indeed tune the evolution of the Hull concavity by adjusting a growth parameter. These results show that the innate material behaviour of the multi-agent model of *P. polycephalum* is suitable for spatially represented problems relating to the definition of shape. In future work we plan to extend the range of geometry and image processing tasks using this approach.

# Funding

This paper was supported by the EU research project "Physarum Chip: Growing Computers from Slime Mould" (FP7 ICT Ref 316366).

# 5 Appendix 1. Culture conditions for experimental growth of *P. polycephalum*

Stock cultures were maintained by cultivating plasmodia of *P. polycephalum* (strain HU554×HU560) on 2% non-nutrient agar in the absence of light at room temperature (22±3$^o$C). Porridge oats (Sainsburys, UK) were supplied as a nutrient substrate and plasmodia were routinely subcultured every 4–5 days, as required.

# 6 Appendix 2. Image Filtering

Filtered images were produced by manually identifying the RGB values of both slime mould and underlying cardboard mask, which were then isolated and re-coloured with a bespoke Processing script. The two separate images were then overlaid in Adobe Photoshop CS6 with the 'difference blend' function. Slime mould is coloured green, cardboard is coloured grey and uncertainty between the two is coloured purple. All other elements of the images are blacked out. These images are included to aid recognition of plasmodial morphology.

# 7 Appendix 3. Particle Model Description

The multi-agent particle approach to modelling *P. polycephalum* uses a population of indirectly coupled mobile particles with very simple behaviours, residing within a 2D diffusive lattice which stores particle positions and the concentration of a generic diffusive factor referred to as chemo-attractant. Collective particle *positions* represent the global pattern of the model plasmodium and collective particle *motion* represents flux within the plasmodium. The particles act independently and iteration of the particle population is performed randomly to avoid any artifacts from sequential ordering. The model is governed by parameters which affect the particle behaviour and the interaction with the environment. For a complete list of parameters see Table 1 which is grouped by category. When a particular parameter is not used (for example in experiments with a fixed population size), the table cell is filled by a dash.

## 7.1 Generation of Emergent Transport Network and Subsequent Morphological Adaptation

The behaviour of the particles occurs in two distinct stages, the sensory stage and the motor stage. In the sensory stage, the particles sample their local environment using three forward biased sensors whose angle from the forwards position (the sensor angle parameter, $SA$), and distance (sensor offset, $SO$) may be parametrically adjusted (Fig. 17a). The offset sensors generate local coupling of sensory inputs and movement to generate the cohesion of the population. The $SO$ distance is measured in pixels and a minimum distance of 3 pixels is required for strong local coupling to occur. The *SO* parameter acts as a scaling parameter, small values result in fine-grained networks whereas larger values result in coarse-grained networks. It was found in [21] that larger *SO* values resulted in the formation of so-called 'vacancy islands' within large blobs of the model plasmodium. For applications where large 'blobs' of model plasmodium are required, these structures can be removed by selecting *SO* from a random value from a pre-set range (see Table 1) for each agent at each scheduler step. During the sensory stage each particle changes its orientation to rotate (via the parameter rotation angle, *RA*) towards the strongest local source of chemo-attractant (Fig. 17b). After the sensory stage, each particle executes the motor stage and attempts to move forwards in its current orientation (an angle from 0–360°) by a single pixel forwards. Each lattice site may only store a single particle and particles deposit chemo-attractant into the lattice only in the event of a successful forwards movement. If the next chosen site is already occupied by another particle the move is abandoned and the particle selects a new randomly chosen direction.

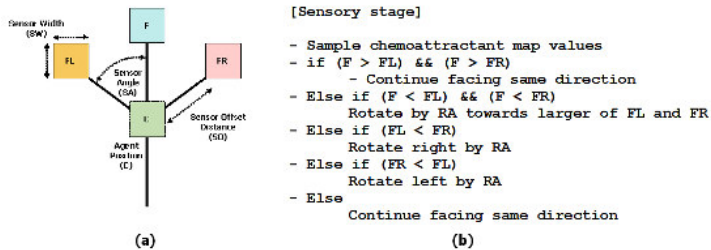

Figure 17: Architecture of a single component of the virtual plasmodium and its basic sensory algorithm. (a) Morphology showing agent position 'C' and offset sensor positions (FL, F, FR), (b) Algorithm for particle sensory stage.

## 7.2 Problem Representation and Experimental Parameters

Source node points were represented by projection of chemo-attractant to the diffusive lattice at their positions indicated by white pixels on the source configuration images. Lattice size varied with each particular arena but varied from 150 pixels minimum size to 400 pixels maximum size. The attractant projection concentration was represented by the $proj_a$ parameter. Projection of repellent sources was implemented by negatively valued projection into the lattice at arena boundary locations causing repulsion of the blob from these regions and is given by $proj_r$. Diffusion in the lattice was implemented at each scheduler step and at every site in the lattice via a simple mean filter of kernel size given by $D_w$. Damping of the diffusion distance, which limits the distance of chemo-attractant gradient diffusion, was achieved by multiplying the mean kernel value by $1 - D_d$ per scheduler step.

### 7.3 Agent Particle Parameters

The model was initialised by creating a population of particles and inoculating the population within the habitable regions of the arena The exact population size differed depending on the particular experiment and is given by the number of particles $p$. Particles were given random initial positions within the habitable area and also random initial orientations. Particle sensor offset distance is given by $SO$. Angle of rotation is given by $RA°$ and sensor angle given by $SA°$. Values of $SA$ and $RA$ differ in experiments depending on whether that experiment reproduced foraging behaviour, or strongly minimising adaptation behaviour. Agent forward displacement was 1 pixel per scheduler step and particles moving forwards successfully deposited chemo-attractant into the diffusive lattice, given by $Dep_t$. This value is less than the attractant projection value $proj_a$, causing the particles to be attracted to projection sites and ultimately constraining the adaptation of the model plasmodium. Both data projection stimuli and agent particle trails were represented by the same chemo-attractant ensuring that the particles were attracted to both data stimuli and other agents' trails.

### 7.4 Representing Light Irradiation Masks

Simulation of light-irradiation masked regions was implemented by damping the sensing of trails by the agents by multiplying by a damping factor between 0 and 1, given by $L_d$ if an agent was within the neighbourhood of a masked area, whose window was given by $L_w$.

### 7.5 Growth and Shrinkage of Model Plasmodium

Adaptation of the blob size was implemented via tests at regular intervals. The frequency at which the growth and shrinkage of the population was executed determined the turnover rate for the population. The frequency of testing for growth was given by the $G_f$ parameter and the frequency for testing for shrinkage is given by the $S_f$ parameter. Growth of the population was implemented as follows: If there were between $G_{min}$ and $G_{max}$ particles in a local neighbourhood (window size given by $G_w$) of a particle, and the particle had moved forward successfully, a new particle was created if there was a space available at a randomly selected empty location in the immediate $3 \times 3$ neighbourhood surrounding the particle.

Shrinkage of the population was implemented as follows: If there were between $S_{min}$ and $S_{max}$ particles in a local neighbourhood (window size given by $S_w$) of a particle the particle survived, otherwise it was deleted. Deletion of a particle left a vacant space at this location which was filled by nearby particles (due to the emergent cohesion effects), thus causing the blob to shrink slightly. As the process continued the model plasmodium shrunk further and adapted its shape to the stimuli provided by the configuration of path source points, arena boundaries and repellent obstacles.

Table 1: Parameters for experiments using the multi-agent model

| Parameter Type | Parameter | Fig | 5a | 5b | 6 | 7 | 8 | 10 | 12 | 13 | 15 | 16 |
|---|---|---|---|---|---|---|---|---|---|---|---|---|
| Agent Parameters | population size | $p$ | 10 | 10 | 800 | 1000 | 3000 | 18000 | 1000 | 1000 | 1000 | 1000 |
| | sensor angle | $SA$ | 22.5 | 22.5 | 45 | 60 | 45 | 60 | 60 | 90 | 90 | 90 |
| | rotation angle | $RA$ | 45 | 45 | 45 | 60 | 45 | 60 | 60 | 45 | 45 | 45 |
| | sensor offset | $SO$ | 5 | 5 | 5 | 5 | 9 | 7 | 13 | 1-19 | 1-19 | 1-19 |
| | agent deposition | $Dep_t$ | 5 | 5 | 15 | 15 | 0.01 | 5 | 5 | 5 | 5 | 5 |
| Diffusion | diffusion window | $D_w$ | 5 | 5 | 3 | 3 | 3 | 3 | 3 | 3 | 3 | 3 |
| | diffusion damping | $D_d$ | 0.1 | 0.1 | 0.1 | 0.1 | 0.07 | 0.05 | 0.1 | 0.05 | 0.05 | 0.05 |
| Stimuli | attractant projection value | $proj_a$ | 12.75 | 12.75 | 127 | - | - | 2.55 | 2.55 | 5 | 5 | 5 |
| | repellent projection value | $proj_r$ | - | - | - | -127 | -127 | - | - | - | - | - |
| Illumination | illumination window | $L_w$ | 3 | 3 | - | - | - | - | - | - | - | - |
| | illumination damping | $L_d$ | 0.9 | 0.9 | - | - | - | - | - | - | - | - |
| Growth | growth frequency | $G_f$ | 3 | 3 | - | - | - | 3 | 5 | 3 | 3 | 3 |
| | growth window | $G_w$ | 9 | 9 | - | - | - | 9 | 9 | 9 | 9 | 9 |
| | growth min | $G_{min}$ | 0 | 0 | - | - | - | 0 | 0 | 0 | 0 | 0 |
| | growth max | $G_{max}$ | 15 | 15 | - | - | - | 20 | 30 | 20 | 20 | 5-30 |
| Shrinkage | Shrinkage frequency | $S_f$ | 3 | 3 | - | - | - | 50 | 50 | 10 | 10 | 10 |
| | shrinkage window | $S_w$ | 5 | 5 | - | - | - | 9 | 9 | 9 | 9 | 9 |
| | shrinkage min | $S_{min}$ | 0 | 0 | - | - | - | 0 | 0 | 0 | 0 | 0 |
| | shrinkage max | $S_{max}$ | 24 | 24 | - | - | - | 80 | 80 | 80 | 80 | 80 |